\documentclass[10pt,superscriptaddress,twocolumn,amsmath,amssymb,aps,prl,showpacs]{revtex4}

\usepackage{mathrsfs}
\usepackage{graphicx}
\usepackage{dcolumn}
\usepackage{bm}
\usepackage{amssymb}
\usepackage{amsmath}
\usepackage{paralist}
\usepackage{color}
\usepackage{ulem}

\usepackage{pdfpages}
\usepackage{subfigure}
\usepackage{cleveref}

\allowdisplaybreaks[4]

\newif\ifNOSUP \NOSUPfalse

\def\be{\begin{equation}}
\def\ee{\end{equation}}
\def\bea{\begin{eqnarray}}
\def\eea{\end{eqnarray}}

\begin{document}

\title{Gr\"uneisen parameters for Lieb-Liniger and Yang-Gaudin models}

\author{Li Peng}
\affiliation{State Key Laboratory of Magnetic Resonance and Atomic and Molecular Physics,
Wuhan Institute of Physics and Mathematics, Chinese Academy of Sciences, Wuhan 430071, China}
\affiliation{University of Chinese Academy of Sciences, Beijing 100049, China}

\author{Yicong Yu}
\email[]{176133205@qq.com}
\affiliation{State Key Laboratory of Magnetic Resonance and Atomic and Molecular Physics,
Wuhan Institute of Physics and Mathematics, Chinese Academy of Sciences, Wuhan 430071, China}

\author{Xi-Wen Guan}
\email[]{xwe105@wipm.ac.cn}
\affiliation{State Key Laboratory of Magnetic Resonance and Atomic and Molecular Physics,
Wuhan Institute of Physics and Mathematics, Chinese Academy of Sciences, Wuhan 430071, China}
\affiliation{Center for Cold Atom Physics, Chinese Academy of Sciences, Wuhan 430071, China}
\affiliation{Department of Theoretical Physics, Research School of Physics and Engineering,
Australian National University, Canberra ACT 0200, Australia}

\date{\today}


\begin{abstract}
%
Using the Bethe ansatz solution, we analytically study  expansionary, magnetic and interacting   Gr\"uneisen parameters (GPs)   for one-dimensional (1D)  Lieb-Liniger and Yang-Gaudin models.
These different  GPs elegantly  quantify   the dependences of  characteristic energy scales   of these quantum gases on the  volume, the magnetic field and the interaction strength, revealing the caloric effects resulted from  the variations  of these potentials. 
 The obtained GPs further  confirm  an   identity which is incurred by   the symmetry of the thermal potential. 
 We also present  universal scaling  behavior of these GPs in the vicinities of  the quantum critical points  driven  by different potentials. 
The divergence of the GPs  not only  provides an experimental identification of non-Fermi liquid nature at quantum criticality but also elegantly   determine low temperature phases of the quantum gases. 
Moreover, the pairing and depairing features in the 1D attractive Fermi gases can be captured by the magnetic and interacting  GPs, facilitating experimental observation of quantum phase transitions. 
 Our results open to further study the  interaction- and magnetic-field-driven quantum refrigeration  and  quantum heat engine in quantum gases of ultracold atoms.

\end{abstract}
\maketitle


\section{I. introduction}

The Gr\"uneisen parameter (GP),  originally introduced to characterize  the frequency  change  due  to the variation of  volume  in  a crystal lattice \cite{Gruneisen_AdP_1908,Gruneisen_AdP_1912}, 
plays an important role in the study of pressure and volume effects in solid state materials. 
Usually, it was  determined from the ratio of thermal expansion to specific heat, quantifying  the  pressure  dependence of characteristic energy scales of solid materials. 
Nowadays, it is  extensively studied in  geophysics \cite{Stacey_PEPI_1995,Shanker_PEPI_2017}, plasma physics \cite{Wang_PRE_2017,Kumar_PJP_2016,Khrapak_PP_2017}, chemistry physics \cite{Mausbach_JCP_2016, Liu_PCCP_2017} and various fields of physics.
The GP  has recently been investigated  in heavy fermion systems \cite{Kuchler_PRL_2003,Kuchler_PRL_2006,Kuchler_STAM_2007}. 
The divergence of the GP  in these systems shows generic signatures of quantum criticality  \cite{Gegenwart_RPP_2016}, revealing the adiabatic magnetocaloric effect of the heavy-fermion metals. 

Dimensionless constants such as the Wilson ratio, 
 Wiedemann-Franz (WF) law \cite{WieF53} and  the Kadowaki-Wood
ratio \cite{KadW86,JacFP09} are very useful in the study of quantum liquids and  electronic transport properties. 
The nature of these ratios essentially reflects the ratio between two types of fluctuations. 
It is well known that the susceptibility(compressibility) Wilson ratio proposed in  \cite{Wilson:1975,Yu_PRB_2016,He_PRB_2017} presents   the ratio between the polarization  (or particle number) fluctuation and the energy fluctuation.  
Although  there have been extensive studies on the GP in various fields of physics,  little work is  carried out  on the GP  for the  quantum gases.
 This is mainly because  both thermal expansion and specific heat in quantum gases\cite{Souza_EJP_2016} are notoriously difficult to be measured in  experimental study.
A recent study \cite{Yu:2019} shows that the magnetocaloric effect (or interaction driven caloric effect) can help to measure the GPs in the controllable systems of ultracold atoms.
The  parameter  \cite{Gruneisen_AdP_1908,Gruneisen_AdP_1912} 
reveals  the  spectrum  change (anharmonicity of the frequency)  to the variation of  the volume of a crystal lattice.
However, in literature,  the formulation of the GP seems to be miscellaneous for  different physical phenomena.
 By the definition of the GP, we first present  an explicit form of the GP in grant canonical ensemble \cite{Yu:2019}
\begin{eqnarray}\label{GCE_GR}
\Gamma= \frac{V\frac{ \mathrm{d}  p}{ \mathrm{d}  T}\vert_{V,N}}{\frac{ \mathrm{d}  E}{ \mathrm{d}  T}\vert_{V,N}}
=\frac{1}{T}\frac{\frac{\partial^2 p}{\partial \mu^2}\frac{\partial p}{\partial T}-\frac{\partial^2 p}{\partial \mu\partial T}\frac{\partial p}{\partial \mu} }{\frac{\partial^2 p}{\partial \mu^2}\frac{\partial^2 p}{\partial T^2}-(\frac{\partial^2 p}{\partial \mu\partial T})^2} 
\end{eqnarray}
where the Maxwell relations and general thermal relation  were used. 
%
%
This expression together with the relation (\ref{GP-1}) presents  a quantitative description of the caloric effect which is induced by  the variation of the system size. 
The magnetocaloric effect is described by  the so-called magnetic GP
\begin{equation}\label{GR_mag}
\Gamma_{\mathrm{mag}}=-\frac{H}{T}\frac{(\partial S/ \partial H)|_{N,T,V}}{(\partial S/\partial T)|_{N,H,V}}=\frac{H}{T}\frac{\partial T}{\partial H}|_{S,N,V},
\end{equation}
where $S$ is the entropy. 
The magnetocaloric effect marks the change of temperature in response to an adiabatic change of $H$, see (\ref{GP-2}). 
This feature  has been used for the  adiabatic demagnetization cooling  \cite{Wolf_PNAS_2011,
Wolf_IJMPB_2014}.
We would like to mention that  in contrast to  some previous studies \cite{Garst_PRB_2005,Ryll_PRB_2014}, here we put the magnetic field $H$ in the numerator to make the GP dimensionless. 
Similarly,  using the Maxwell relations and general thermal relations, we can obtain the expression of the magnetic GP  in grant canonical ensemble \cite{Yu:2019}
\begin{eqnarray}\label{GCE_GRmag}
\Gamma_{\mathrm{mag}}=-\frac{H}{T}\frac{\frac{\partial^2 p}{\partial \mu^2}\frac{\partial^2 p}{\partial H\partial T}-\frac{\partial^2 p}{\partial \mu\partial H}\frac{\partial^2 p}{\partial \mu \partial T} }{\frac{\partial^2 p}{\partial \mu^2}\frac{\partial^2 p}{\partial T^2}-(\frac{\partial^2 p}{\partial \mu\partial T})^2}.
\end{eqnarray}

In ultracold atoms the interaction between two colliding atoms can be tuned  by  Feshbach resonances or confinement-induced resonances. 
The contact interaction can be regarded as a potential which changes the free energy in grand canonical ensemble. The interacting caloric effect in quantum gases has  been rarely  studied \cite{Yu:2019,Chen:2019}.
 Similarly, the interacting GP, describing  the change of temperature in response to an adiabatic change of the  interaction, can be  given by  
\begin{eqnarray}
\label{GCE_GRint}
\Gamma_{\mathrm{int}}=-\frac{c\frac{\partial S}{\partial c}\vert_{N,H,T,V}}{T\frac{\partial S}{\partial T}\vert_{N,H,{ c},V}}
=-\frac{c}{T}\frac{\frac{\partial^2 p}{\partial \mu^2}\frac{\partial^2 p}{\partial c\partial T}-\frac{\partial^2 p}{\partial \mu\partial c}\frac{\partial^2 p}{\partial \mu \partial T} }{\frac{\partial^2 p}{\partial \mu^2}\frac{\partial^2 p}{\partial T^2}-(\frac{\partial^2 p}{\partial \mu\partial T})^2},
\end{eqnarray}
%
where $c$ denotes the interaction strength between atoms.
To summarize, the relations between the caloric effect and  the GPs are given by 
\begin{eqnarray}
\frac{\partial T}{\partial V} \mid_{S,N,H,c}&=&\frac{T}{V} \Gamma, \label{GP-1} \\
\frac{\partial T}{\partial H} \mid_{S,N,V,c}&=&\frac{T}{H} \Gamma_{\text{mag}}, \label{GP-2}\\
\frac{\partial T}{\partial c} \mid_{S,N,V,H}&=&\frac{T}{c} \Gamma_{\text{int}}  \label{GP-3}
\end{eqnarray}
that provide  plausible  experimental measurements of the  GPs caused by the changes of the volume, magnetic field and interaction, respectively. 
The given relation (\ref{GP-3}) marks the change of temperature in response to an adiabatic change of the interaction strength. 
Similar to the magnetic GP, the interacting  GP  is   experimentally measurable and  can  be used to quantify  the  adiabatic refrigeration driven by interaction. 
We will study these three types of the GPs   for the 1D Bose and Fermi gases in the following sections.

In this paper, we systematically  study  the GPs  for 1D quantum gases using Bethe ansatz solutions.
We analytically obtain  the system size,  magnetic field and interaction-driven  GPs for the  Lieb-Liniger model   \cite{Lieb_PR_1963} and Yang-Gaudin model \cite{Yang_PRL_1967,Gaudin:1967}. 
 The obtained GPs show volume,  magnetic field and interaction dependences of  characteristic energy scales   of the quantum gases and present  the caloric effects resulted from the variations  of these potentials. 
Our results confirm the   identity  among the three types of GPs \cite{Yu:2019} in 1D quantum systems
\begin{align}
d\cdot \Gamma +2\Gamma_{\mathrm{mag}} + \Gamma_{\mathrm{int}}=2,
\label{eq:identity}
\end{align}
here $d=1$ is the dimensionality. 
 We also derive  the universal singular behaviors of the GPs at the quantum critical points in the attractive Yang-Gaudin model  \cite{Yang_PRL_1967,Gaudin:1967},  driven either by the magnetic field or the interaction. 
 The divergent behaviors of  the GPs near quantum critical points  show the enhancement of the caloric effects at phase boundaries and help to elegantly obtain the phase diagram and to realize quantum cooling.

The outline of this paper is as follows. 
In section II, we study the GPs for the 1D Lieb-Liniger Bose gas. Analytical expression of the interacting  GP is presented. 
In section III, we present exact analytical  expressions of the GPs for the Yang-Gaudin model   in the fully polarized phase, fully paired phase and 
the Fulde-Ferrell-Larkin-Ovchinnikov (FFLO) like  pairing phase. 
We also studied the phase transitions of  the model in terms of different GPs. 
We conclude with a brief summary in section IV.

\section{II. Gr\"{u}neisen parameter for the  Lieb-Liniger model }
\label{sectionIII}

The  Lieb-Liniger model \cite{Lieb_PR_1963}, which describes the 1D interacting bosons, is a prototypical Bethe ansatz solvable model \cite{korepin1997quantum}. 
It is one of the most extensively studied many-body systems in ultracold atoms. 
The Hamiltonian of the model is ($\hbar=2m=1$)
\begin{equation}
  \hat{H} = - \sum_{i = 1}^N \frac{\partial_{}^2}{\partial {x_i}^2} + 2 c \sum_{i
  < j}^N \delta (x_i - x_j), \label{Ham}
\end{equation}
where  $N$ is the total number of the spinless bosons constrained by periodic boundary conditions on a line of length $L$.
For repulsive (attractive) contact interaction, the interaction strength $c>0$ ($c<0$). 
The  coupling constant $c = -2 \hbar^2/m a_{\rm 1D}$  is determined  by the  1D scattering length, given by $a_{1D}=\left( -a_{\perp }^{2}/2a_{s}\right) \left[ 1-C\left(
a_{s}/a_{\perp }\right) \right]$\cite{Olshanii_PRL_1998,Dunjko_PRL_2001,Olshanii_PRL_2003}. 
Here the numerical constant $C\approx 1.4603$.

The Bethe ansatz wave function for the Lieb-Liniger model  (\ref{Ham}) is given by 
\begin{equation}
\Psi= \sum_p(-1)^p\Big[\prod_{1\le i<j\le N}\Big(1+\frac{\mathrm{i} k_{p_j}-\mathrm{i}  k_{p_i}}{c}\Big)\Big]\exp\Big(\sum_{j=1}^N\mathrm{i} k_{p_j}x_j\Big)\label{wave-function}
 \end{equation}
where  $p$ stands for $N!$  permutations  of integers $1,2,...,N$. The pseudo momenta $k_j$ satisfies the following BA equation
\begin{equation}
  \mathrm{e}^{\text{i} k_j L} = - \prod_{l = 1}^N \frac{k_j - k_l + \text{i} c}{k_j - k_l
  - \text{i} c}.\label{BAE}
\end{equation}
The solution to the Eq.(\ref{BAE}) provides complete spectra of the Lieb-Liniger model with $ E=\sum_{j}^N k^2_j$. 

In 1969, Yang and Yang \cite{yang1969thermodynamics} introduced the particle hole ensemble to describe the thermodynamics of the model in equilibrium, which is  later called the thermodynamics Bethe ansatz (TBA) approach \cite{Takahashi_2005}. 
In terms of the dressed energy $\varepsilon(k)=T\ln (\rho^h(k)/\rho(k))$ defined with respect to the quasi-momentum $k$ at finite temperature $T$, the TBA equation is given by 
\begin{equation}\label{TBA}
  \varepsilon(k) = k^2 - \mu - \frac{T c}{\pi} \int_{- \infty}^{+ \infty} \frac{{\rm d}q}{c^2 + (k -
  q)^2} \ln \left( 1 + \mathrm{e}^{- \frac{\varepsilon(q)}{T}} \right),
\end{equation}
where $\mu$ is the chemical potential. The dressed energy $\varepsilon(k)$ plays the role of excitation energy measured from the energy level $\varepsilon(k_F)=0$, where $k_F$ is the Fermi-like momentum. 
The pressure $p$ is given in terms of the temperature and chemical potential 
\begin{equation}\label{definition_pressure}
  p = \frac{T}{2 \pi} \int^{+ \infty}_{- \infty} \ln \left( 1 + \mathrm{e}^{-
  \frac{\varepsilon(k)}{T}} \right) {\rm d}k,
\end{equation}
serving as the equation of  states of the system.

\begin{figure}[]
\begin{center}
\includegraphics[width=0.9\linewidth]{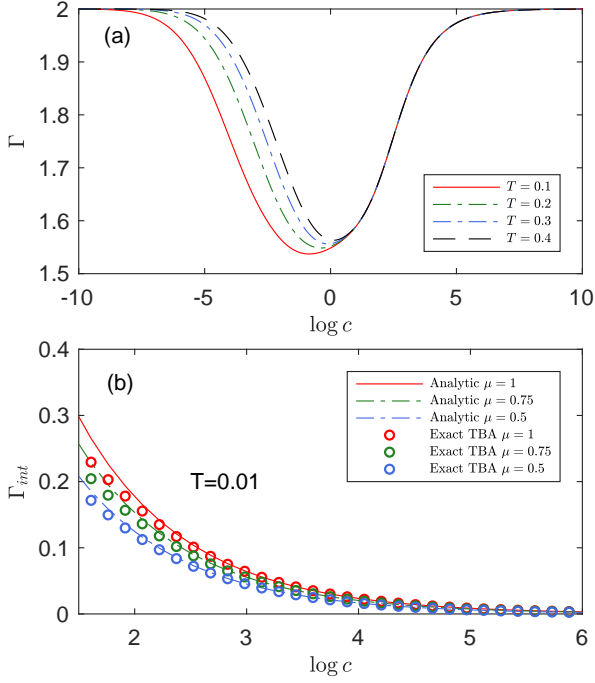}
\end{center}
\caption{(a) The Gr\"uneisen parameter (\ref{GCE_GR}) versa interaction strength $\log c$ for the 1D Bose gas  at different temperatures. Here the total density is fixed $n=1$.  In the weakly coupling limit, the system behaves as a free Bose gas for which the GP $\Gamma =2$.  Whereas in the strongly coupling limit, it gives a ferminized  Tonks-Girardeau gas with  $\Gamma =2$. This value of the GP shows a scaling invariant nature. 
(b) The interacting  Gr\"uneisen parameter (\ref{GCE_GRint}) versa interaction strength $\log c$ for  the strong coupling regime. Here we set $T=0.01$ and $\mu=1,\,0.75,\, 0.5$.  For a strong repulsion, we observe a good agreement  between the numerics  (asterisks) and analytical (solid lines) result Eq. (\ref{gp_int-strong}). 
}\label{Fig1}
\end{figure}

For a strong repulsive interaction,  the gas is called the Tonks-Girardeau (TG) gas \cite{girardeau1960}. 
In the TG limit,  the Bose-Fermi mapping can be used to study  the ground-state properties of the Bose gas, where the wave function of bosons can be written as  the product of the sign function and the wave function of the non-interacting fermions. 
From the expressions of the GPs (\ref{GCE_GR}) and (\ref{GCE_GRint}), by solving the TBA equation (\ref{TBA}) we can numerically obtain the  volume and interacting GPs, see Fig.{\ref{Fig1}}.
In Fig.~{\ref{Fig1}}(a) we see  a subtle change of the GP as the interaction strenth varies from zero to infinity. 
In these two  limits, the scaling invariance leads to $\Gamma =2$ in 1D.

It is particularly interesting to investigate the interacting  GP (\ref{GCE_GRint}). 
In the strong interaction regime, $\gamma=c/n\gg1$, $T/c^2\ll1$,  the TBA Eq.(\ref{TBA}) can be expanded in the powers of $1/c$ (up to the order of $O(1/c^3)$) \cite{Guan_JPA_2011}
\begin{equation}\label{strong_TBA}
	\varepsilon(k)\approx k^2 -\mu-\frac{2c}{c^2+k^2}p-\frac{1}{2\sqrt{\pi}c^3}T^{\frac{5}{2}}\text{Li}_{\frac{5}{2}}(-\mathrm{e}^{\frac{A_0}{T}}),
\end{equation}
where $
A_0=\mu+\frac{2p}{c}-\frac{4\mu^{5/2}}{15\pi|c|^3}$ 
and $\text{Li}_ {s}(z)=\frac{1}{\Gamma(s)}\int_{0}^{+ \infty}\frac{t^{s-1}}{\mathrm{e}^t/z-1}{\rm d} t$ denotes  the polylogarimic  function.
For simplicity, here we only consider the calculation up to  the order of $1/c^2$. 
Substituting the  Eq.(\ref{strong_TBA}) into the Eq.(\ref{definition_pressure}), we  obtain the pressure 
\begin{eqnarray}\label{pressure}
  p &=& -\frac{1}{2\sqrt{\pi}}T^{\frac{3}{2}} \text{Li}_{\frac{3}{2}}(-\mathrm{e}^{\frac{A}{T}})
\end{eqnarray}
with $A = \mu+\frac{2}{c}p$.
In the follow calculation, we further restrict to  the low temperature regime  $T\ll c^2$ in order to  give an explicit form  of the Gr\"uneisen parameter in the strong coupling.
Using  the series expansion of the polylogarimic  function \cite{mohankumar2007two}
\begin{eqnarray}\label{polylog_function}
  \text{Li}_s(-\mathrm{e}^x)=-\frac{x^s}{\Gamma(s+1)}[1+\frac{\Gamma(s+1)}{6\Gamma(s-1)}(\frac{\pi}{x})^2+\cdots],
\end{eqnarray}
we further get the equation of the state 
\begin{eqnarray}\label{pressure_chemical}
  p&=& \frac{2}{3\pi}\mu^{\frac{3}{2}}+\frac{4}{3\pi^2c}\mu^2+\frac{28}{9\pi^3c^2}\mu^{\frac{5}{2}}+O(\frac{1}{c^3}),
\end{eqnarray}
then $A=\mu+\frac{4}{3\pi c}\mu^{\frac{3}{2}}+\frac{8}{3\pi^2c^2}\mu^2+O(\frac{1}{c^3})$.
The pressure essentially  depends on the chemical potential and interaction.


Substituting  Eq.(\ref{pressure_chemical})  into (\ref{pressure}) and by iteration, after length  algebraic calculation, we obtain the following derivatives of pressure with respect to chemical potential and temperature
\begin{eqnarray}
  \frac{\partial p}{\partial \mu}&=&\frac{1}{\pi}\mu^{\frac{1}{2}}+\frac{8}{3c\pi^2}\mu+\frac{70}{9c^2\pi^3}\mu^{\frac{3}{2}}+O(\frac{1}{c^3})
\nonumber\\
\frac{\partial^2 p}{\partial \mu^2}&=&\frac{1}{2\pi}\mu^{-\frac{1}{2}}+\frac{8}{3c\pi^2}+\frac{35}{3c^2\pi^3}\mu^{\frac{1}{2}}+O(\frac{1}{c^3})\nonumber\\
  \frac{\partial p}{\partial T}&=&T(\frac{\pi}{6}\mu^{-\frac{1}{2}}+\frac{2}{9c}+\frac{5}{9\pi c^2}\mu^{\frac{1}{2}})+O(\frac{1}{c^3})\nonumber\\
  \frac{\partial^2 p}{\partial \mu \partial T}&=&T(-\frac{\pi}{12}\mu^{-\frac{3}{2}}+\frac{5}{18\pi c^2}\mu^{-\frac{1}{2}})+O(\frac{1}{c^3})\nonumber\\
  \frac{\partial^2 p}{\partial T^2}&=&\frac{\pi}{6}\mu^{-\frac{1}{2}}+\frac{2}{9c}+\frac{5}{9\pi c^2}\mu^{\frac{1}{2}}+O(\frac{1}{c^3}).
\end{eqnarray}
By  the definition of the GP (\ref{GCE_GR}) and (\ref{GCE_GRint}), using the above derivatives,  we obtain the explicit expression 
\begin{eqnarray}\label{gp-strong}
  \Gamma=2-\frac{4\sqrt{\mu}}{\pi c}-
  \frac{8\mu}{3\pi^2 c^2}+O(\frac{1}{c^3}), 
\end{eqnarray}
and
\begin{eqnarray}\label{gp_int-strong}
  \Gamma_{\mathrm{int}}=\frac{4\sqrt{\mu}}{\pi c}+
  \frac{8\mu}{3\pi^2 c^2}+O(\frac{1}{c^3}).
\end{eqnarray}
Fig.{\ref{Fig1}} (b) 
 shows that  the analytical result  (\ref{gp_int-strong}) agrees well with the numerical calculation. 
This clearly indicates a strong repulsion drives the system into a deal gas with a scaling homogeneous spectrum in the limit $c\to \infty$. 
The two figures in Fig.{\ref{Fig1}} clearly indicate different features of the expansionary and interacting GPs. 
Our result further confirms the universal identity (\ref{eq:identity}), namely
\begin{align}
\Gamma + \Gamma_{\mathrm{int}}=2. 
\end{align}
This relation can be used to investigate interaction effect in quantum gases, see the analysis on quantum heat engine in 1D Bose gas \cite{Chen:2019}.

\section{III. Gr\"{u}neisen parameter for the Yang-Gaudin model }

%
In this section we will  study the  magnetic and interacting GPs for the 1D  Yang-Gaudin model  with an attractive interaction \cite{Yang_PRL_1967,Gaudin:1967}. 
The existence of a Fulde-Ferrell-Larkin-Ovchinnikov (FFLO) pairing state in the  interacting Fermi gas have been predicted  by exact solutions \cite{Orso:2007,Hu:2007,Guan:2007}. 
Recent breakthrough experiments on trapped ultracold fermionic atoms confined to 1D  have provided a deep  understanding of such a novel FFLO phase of  the Yang-Gaudin model \cite{Liao_Nature_2010}, see a review  \cite{Guan_RMP_2013}.
Here we will study the magnetic and interacting  GPs throughout the full phase diagram of the model. 
 We will show universal  divergent feature of the GPs near  quantum phase transitions,  which can be used to  probe  quantum scaling, caloric effects and quantum refrigeration  in quantum  many-body systems.

The Yang-Gaudin model \cite{Yang_PRL_1967,Gaudin:1967} described the  1D  spin-$\frac{1}{2}$ Fermi gas with a $\delta$-function interaction.  The Hamiltonian reads
\begin{equation}
\hat{H} = - \sum_{i = 1}^N \frac{\partial_{}^2}{\partial {x_i}^2} + 2 c \sum_{i <
j}^N \delta (x_i - x_j)-\frac{1}{2}H(N_{\downarrow}-N_{\uparrow}).
\end{equation}
in which the terms are the kinetic energy, interaction energy and Zeeman energy, respectively. Here $N$ is the total number of fermions in a length $L$ and $c<0$ indicate the contact attractive interaction.
The BA equations were given by  \cite{Yang_PRL_1967,Gaudin:1967}
\begin{eqnarray}
\mathrm{e}^{\text{i}k_jL} &=& \prod_{\alpha=1}^M\frac{k_j-\Lambda_\alpha+\text{i}c/2}{
k_j-\Lambda_\alpha-\text{i}c/2}, \nonumber\\
\prod_{j=1}^N\frac{\Lambda_\alpha- k_j+\text{i}c/2}{\Lambda_\alpha -k_j-\text{i}c/2} &=&
-\prod_{\beta=1}^M\frac{\Lambda_\alpha-\Lambda_\beta+\text{i}c}{\Lambda_\alpha-
\Lambda_\beta-\text{i}c}.
\end{eqnarray}
The energy of the system is given by $ E=\sum_{j}^N k^2_j$.
The model has spin population imbalance caused by a difference in the number of spin-up and spin-down atoms. 
The key features of this ground state  phase diagram were experimentally confirmed using finite temperature density profiles of trapped fermionic ${}^6$Li atoms  \cite{Liao_Nature_2010}, where three quantum phases fully paired state, partially polarized FFLO like state and fully-polarized state exist in chemical potential-effective magnetic field plane, see \cite{Guan_RMP_2013,Yu_PRB_2016,Guan_PRA_2011}.

In order to build up the thermodynamic Bethe ansatz approach to the 1D attractive Fermi gas, 
we  define  the dressed energy as $\epsilon^{b}(k)=T \ln (\rho_2^h(k)/ \rho_2(k))$ and $\epsilon^{u}(k)=T \ln (\rho_1^h(k)/ \rho_1(k))$,  corresponding to paired fermions and unpaired fermions in the grand canonical ensemble,  respectively.
According to the Yang-Yang method  \cite{yang1969thermodynamics},  by  minimizing  the free energy   the TBA equations are given by \cite{Takahashi_2005,Guan_RMP_2013}

\begin{eqnarray}
\epsilon^{b}(k)&=&2(k^2-\mu-\frac{1}{4}c^2)+Ta_2*\ln(1+\mathrm{e}^{%
-\epsilon^{b}(k)/T})\nonumber\\
   &&+Ta_1*\ln(1+\mathrm{e}^{-\epsilon^{u}(k)/T}), \nonumber\\
\epsilon^{u}(k)&=&k^2-\mu-\frac{1}{2}H+Ta_1*\ln(1+\mathrm{e}^{-\epsilon^{b}(k)/T})\nonumber\\
   &&-T\sum^{\infty}_{l=1}a_l*\ln(1+\eta_l^{-1}(k)), \nonumber\\
\ln \eta_{l}(\lambda)&=&\frac{lH}{T}+a_l*\ln(1+\mathrm{e}^{-\epsilon^u(\lambda)/T})\nonumber\\
   &&+\sum^{\infty}_{m=1}T_{lm}*\ln(1+\eta_{m}^{-1}(\lambda)),\label{TBA-F}
\end{eqnarray}
where $a_n(x)=\frac{1}{2\pi}\frac{n\vert{c}\vert}{(\frac{n\vert{c}\vert}{2})^2+x^2}$ and the function $T_{lm}$ is given in \cite{Guan_RMP_2013}. In the above equations,  `$*$' denotes convolution, namely $ a*f(x)=\int a(x-y)f(y){\rm d}y$. And the $\eta_{l}(\lambda):\equiv \xi^h_l(\lambda)/\xi_l(\lambda)$ are the ratio between  the hole density $\xi^h_l(\lambda)$  and the particle density $\xi_l(\lambda)$ of the length-$l$ strings, associated with  the excitations of magnons. The pressure  is given by
\begin{eqnarray}
p&=&\frac{T}{\pi}\int_{-\infty}^{\infty}\mathrm{d}k\ln(1+\mathrm{e}^{-
\epsilon^{b}(k)/T})\nonumber\\
  &&+\frac{T}{2\pi}\int_{-\infty}^{\infty}\mathrm{d}
k\ln(1+\mathrm{e}^{-\epsilon^{u}(k)/T}).\label{TBA-pressture}
\end{eqnarray}
The TBA equations (\ref{TBA-F}) provide us with an analytical way to  study  the Gr\"uneisen parameters. 

By taking integration by part in the pressure (\ref{TBA-pressture}), we then obtain the pressure $p= {\widetilde{p}}^{(1)}+ {\widetilde{p}}^{(2)} $ in terms of  the leading terms of the  effective pressures 
\begin{eqnarray}
  {\widetilde{p}}^{(1)} &=& -\frac{1}{2\sqrt{2\pi}}t^{\frac{3}{2}} \text{Li}_{\frac{3}{2}}(-\mathrm{e}^{{\widetilde{A}}^{(1)}/t}),\nonumber\\
  {\widetilde{p}}^{(2)} &=& -\frac{1}{2\sqrt{\pi}}t^{\frac{3}{2}} \text{Li}_{\frac{3}{2}}(-\mathrm{e}^{{\widetilde{A}}^{(2)}/t}),
 \end{eqnarray}
where $ {\widetilde{A}}^{(1)} = {\widetilde{A}}_{0}^{(1)}-2{\widetilde{p}}^{(2)}$ and 
$  {\widetilde{A}}^{(2)} = {\widetilde{A}}_{0}^{(2)}-{\widetilde{p}}^{(2)}-4{\widetilde{p}}^{(1)}$. 
The superscripts $(1)$ and $(2)$  denote unpaired and paired fermions, respectively. 
Here we denote   the effective chemical potentials
$  {\widetilde{\mu}}_1=\widetilde{\mu}+\frac{1}{2}h$ and 
 $ {\widetilde{\mu}}_2=2\widetilde{\mu}+1$ for  unpaired fermions and paired fermions, respectively. 
 For convenience in our later calculation, we will  use the following dimensionless quantities 
\begin{eqnarray*}
\widetilde{A}^{(r)}&=&\frac{A^{(r)}}{\frac{1}{2}{\lvert c\rvert}^2}, \qquad
\widetilde{\mu}=\frac{\mu}{\frac{1}{2}{\lvert c\rvert}^2}, \qquad h=\frac{H}{%
\frac{1}{2}{\lvert c\rvert}^2}, \\
\qquad t&=&\frac{T}{\frac{1}{2}{\lvert c\rvert}%
^2}, \qquad \widetilde{p}^{(r)}=\frac{p^{(r)}}{\frac{1}{2}{\lvert c\rvert}^3}. 
\end{eqnarray*}

\begin{widetext}

\begin{figure}[t]
\begin{center}
\includegraphics[width=0.8\linewidth]{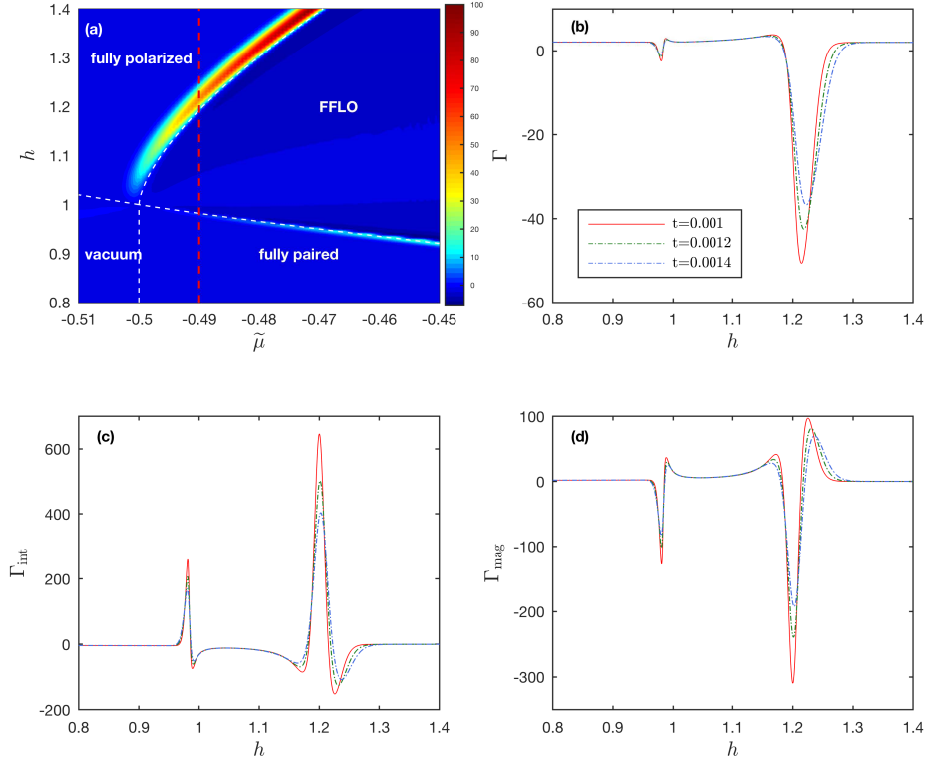}
\end{center}
\caption{ (a) Phase diagram is mapped out from the contour plot  of volume-driven GP   Eq.(\ref{GCE_GR})  at  $t=0.001$.  The values of the interaction GP have  sudden enhancement near the phase boundaries, showing the quantum phase transitions between two different phases among fully-paired, fully-polarized and FFLO like phases. The white dashed lines stand for the analytical zero temperature critical fields, see \cite{Guan_PRA_2011}.  (b)-(d)  show the volume, interaction and magnetic field    driven GPs at $\tilde{\mu}=-0.49$ and  $t=0.001, 0.0012, 0.0014$ respectively,   see Eqs.(\ref{GCE_GR}), (\ref{GCE_GRint}) and (\ref{GCE_GRmag}). These GPs characterize universal scaling behaviour  near a quantum phase transition.  }
\label{Fig2}
\end{figure}

\end{widetext}

In the Fig.~\ref{Fig2}(a), we contour plot the GP Eq.(\ref{GCE_GR}), which shows a full phase diagram at low temperature and divergent behaviors near the phase boundaries. At the phase boundaries, these GPs display  universal divergent behaviors.   Fig.\ref{Fig2} (b)-(d) present different GPs for a fixed the effective chemical potential $\tilde{\mu}=-0.49$ and  temperature $t=0.001$, $0.0012$, $0.0014$, respectively, showing  sharp   peaks near the phase transition between the 
FFLO like phase and fully polarized phase as well as between the fully paired phase and the FFLO like state.  At these  phase transitions, the GPs  show a strong caloric effect due to the entropy accumulation, facilitating realization of quantum cooling and quantum heat engine. What follows will derive  the GPs for different phases and  scaling functions  of the GP at the quantum criticality. 
 
\subsection{III.1. The GPs for the fully polarized phase}
 For a non-interacting system, scaling invariance gives rise to a constant value of the GP, i.e. $\Gamma =2/d$, here $d$ is the dimensionality. 
For the Yang-Gaudin model, in the fully polarized phase, we have ${\widetilde{A}}^{(1)}>0$ and ${\widetilde{A}}^{(2)}<0$.
 Therefore  only the pressure of the fully polarized fermions ${\widetilde{p}}^{(1)}$ contributes to the total pressure at low temperatures, namely 
\begin{eqnarray}
  {\widetilde{p}}^{(1)} &=& -\frac{1}{2\sqrt{2\pi}}t^{\frac{3}{2}} \text{Li}_{\frac{3}{2}}(-\mathrm{e}^{{\widetilde{A}}^{(1)}/t}),\nonumber\\
  {\widetilde{A}}^{(1)} &=& {\widetilde{A}}_{0}^{(1)}.
\end{eqnarray}
We may  analytically obtain a series of all the derivatives of the pressure with respect to the temperature and chemical potential in the limit $t\ll{\widetilde{A}}_{0}^{(1)}\ll1$
\begin{eqnarray}
  \widetilde{p}&=&\frac{\sqrt{2}}{3\pi}{{\widetilde \mu}_1}^\frac{3}{2}+O({\widetilde \mu}_1^2),\nonumber\\
  \frac{\partial \widetilde{p}}{\partial \widetilde \mu}&=&\frac{1}{\sqrt{2}\pi}{{\widetilde \mu}_1}^\frac{1}{2}+O({\widetilde \mu}_1),\nonumber\\
  \frac{\partial^2 \widetilde{p}}{\partial {\widetilde \mu}^2}&=&\frac{1}{2\sqrt{2}\pi}{{\widetilde \mu}_1}^{-\frac{1}{2}}+O(1),\nonumber\\
  \frac{\partial \widetilde{p}}{\partial t}&=&t(\frac{\pi}{6\sqrt{2}}{{\widetilde \mu}_1}^{-\frac{1}{2}}+O(1)),\nonumber\\
  \frac{\partial^2 \widetilde{p}}{\partial t^2}&=&\frac{\pi}{6\sqrt{2}}{{\widetilde \mu}_1}^{-\frac{1}{2}}+O(1), \nonumber\\
  \frac{\partial^2 \widetilde{p}}{\partial \widetilde \mu\partial t}&=&t(-\frac{\pi}{12\sqrt{2}}{{\widetilde \mu}_1}^{-\frac{3}{2}}+O({\widetilde \mu_1}^{-\frac{1}{2}})), \nonumber\\
    \frac{\partial^2 \widetilde{p}}{\partial h\partial  t}&=&-t(\frac{\pi}{24\sqrt{2}}{{\widetilde \mu}_1}^{-\frac{3}{2}}+O({\widetilde \mu_1}^{-\frac{1}{2}})),\nonumber\\
  \frac{\partial^2 \widetilde{p}}{\partial {\widetilde \mu}\partial  h}&=&\frac{1}{4\sqrt{2}\pi}{{\widetilde \mu}_1}^{-\frac{1}{2}}+O(1).
\end{eqnarray}
It is straightforward to obtain the Gr\"{u}neisen parameter $\Gamma \approx 2$. While, 
with the help of the definition Eq.(\ref{GCE_GRmag}), the magnetic GP  is given by $ \Gamma_{\mathrm{mag}}=0$.

\subsection{III.2. The GPs for the  fully paired phase}
In the fully paired phase, the system consists of pure pairs in the ground state. At low temperatures, 
we have ${\widetilde{A}}^{(1)}<0$ and ${\widetilde{A}}^{(2)}>0$. Therefore, the effective pressure of paired fermions   ${\widetilde{p}}^{(2)}$ mainly contributes to the total pressure.  It follows that 
\begin{eqnarray*}
  {\widetilde{p}}^{(2)} &=& -\frac{1}{2\sqrt{\pi}}t^{\frac{3}{2}} \text{Li}_{\frac{3}{2}}(-\mathrm{e}^{{\widetilde{A}}^{(2)}/t}), \\
  {\widetilde{A}}^{(2)} &=& {\widetilde{A}}_{0}^{(2)}-{\widetilde{p}}^{(2)}. 
\end{eqnarray*}
Similarly, we obtain the leading order of the derivatives with respect to chemical potential and temperature
\begin{eqnarray}
  \widetilde{p}&=&\frac{2}{3\pi}{{\widetilde{\mu}}_2}^\frac{3}{2}-\frac{2}{3\pi^2}{{\widetilde{\mu}}_2}^2
  +\frac{7}{9\pi^3}{{\widetilde{\mu}}_2}^\frac{5}{2}+O({{\widetilde{\mu}}_2}^3),\nonumber\\
  \frac{\partial \widetilde{p}}{\partial \widetilde \mu}&=&\frac{2}{\pi}{{\widetilde{\mu}}_2}^\frac{1}{2}
  -\frac{8}{3\pi^2}{{\widetilde{\mu}}_2}+
  \frac{35}{9\pi^3}{{\widetilde{\mu}}_2}^\frac{3}{2}+O({{\widetilde{\mu}}_2}^2),\nonumber\\
  \frac{\partial^2 \widetilde{p}}{\partial {\widetilde \mu}^2}&=&
   \frac{2}{\pi}{{\widetilde{\mu}}_2}^{-\frac{1}{2}}
   -\frac{16}{3\pi^2}+\frac{35}{3\pi^3}{{\widetilde{\mu}}_2}^{\frac{1}{2}}+O({{\widetilde{\mu}}_2}), \nonumber\\
  \frac{\partial \widetilde{p}}{\partial t}&=&\frac{\pi t}{6}({{\widetilde{\mu}}_2}^{-\frac{1}{2}}-\frac{2}{3\pi}+\frac{5}{6\pi^2}{{\widetilde{\mu}}_2}^{\frac{1}{2}}+O({{\widetilde{\mu}}_2})), \nonumber\\
  \frac{\partial^2 \widetilde{p}}{\partial t^2}&=&\frac{\pi}{6}({{\widetilde{\mu}}_2}^{-\frac{1}{2}}-\frac{2}{3\pi}+\frac{5}{6\pi^2}{{\widetilde{\mu}}_2}^{\frac{1}{2}}+O({{\widetilde{\mu}}_2})), \nonumber\\
  \frac{\partial^2 \widetilde{p}}{\partial \widetilde \mu\partial t}&=&t(-\frac{\pi}{6}{{\widetilde{\mu}}_2}^{-\frac{3}{2}}+\frac{5}{36\pi}{{\widetilde{\mu}}_2}^{-\frac{1}{2}}+O(1)), \nonumber\\
   \frac{\partial^2 \widetilde{p}}{\partial h\partial  t}&=&0, \nonumber\\
  \frac{\partial^2 \widetilde{p}}{\partial {\widetilde \mu}\partial  h}&=&0.
\end{eqnarray}
Substituting  these derivatives into  Eq.(\ref{GCE_GR}), we obtain the GP 
\begin{equation}\label{Fermi_P}
\Gamma=2+\frac{2\sqrt{{\widetilde{\mu}}_2}}{\pi}-\frac{2{\widetilde{\mu}}_2}{3\pi^2}. 
\end{equation}
The last two terms in this expression indicate the interaction effect of the spin singlet interaction. 
In fact, for the fully paired phase, the strongly attractive Fermi gas without polarization can be regarded as the  super Tonks-Girardeau gas composed of bosonic Fermi pairs with a weakly attractive pair-pair interaction. By a transformation $c\rightarrow-c$, $\mu_2=2\mu$, we prove that Eq.(\ref{gp-strong}) and  Eq.(\ref{Fermi_P}) are equivalent, i.e. 
\begin{eqnarray}
\Gamma &=&2-\frac{4\sqrt{\mu}}{\pi c}-\frac{8\mu}{3\pi^2 c^2}.
\end{eqnarray}
Thus we see that the GP for the attractive Fermionic  pairs is equivalent to  that of  the hard-core bosons, see (\ref{gp-strong}). 
In addition, the magnetic GP (\ref{GCE_GRmag}) can be calculated  $  \Gamma_{\mathrm{mag}}=0$. 

\subsection{III.3. The GPs for the FFLO like pairing phase}

\begin{figure}[]
\begin{center}
\includegraphics[width=1.0\linewidth]{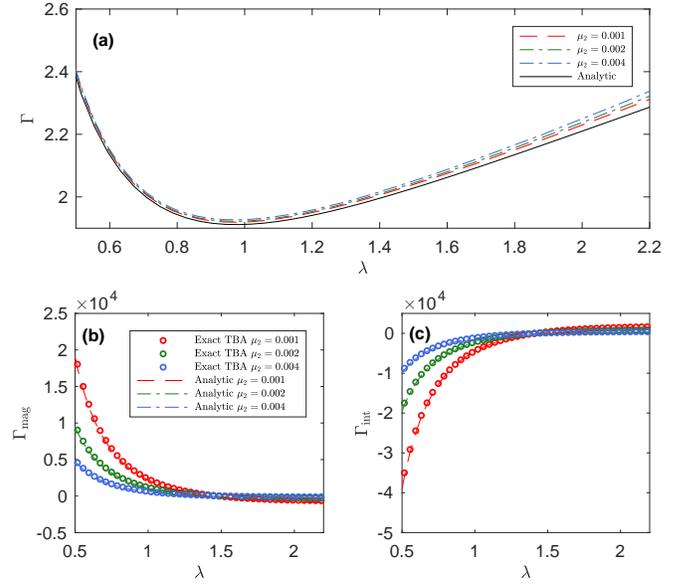}
\end{center}
\caption{(a) The Gr\"uneisen parameter Eq.(\ref{GCE_GR}) vs the effective chemical potential ratio $\lambda$ for  the FFLO-like phase. 
 Dashed lines show the numerical result from the Eq.(\ref{GCE_GR}) and the TBA equations (\ref{TBA-F}). The black solid line present the result of Eq.(\ref{gr_FFLO}) have a good agreement with the numerical result.  (b) The magnetic Gr\"uneisen parameter Eq.(\ref{GCE_GRmag}) vs the effective chemical potential ratio $\lambda$ for  the FFLO-like phase.  (c) The interaction Gr\"uneisen parameter Eq.(\ref{GCE_GRint}) vs the effective chemical potential ratio $\lambda$ for  the FFLO-like phase. A good agreement  between the numerical (symbols)  and analytical (dashed lines) results in the low density and low temperature limits.
 All shown data have been set for low temperature $T=1.0\times 10^{-5}$, $|c|=10 $ in strong coupling region  and the chemical potential $\mu_2=0.001,\,0.002,\, 0.004$, respectively. }
\label{Fig3}
\end{figure}

In  the FFLO like pairing phase, the system exhibits two states: paired and unpaired  fermions. The spectra of the system are  subtly influenced  by the system size, interaction strength and magnetic field.  In this phase, ${\widetilde{A}}^{(1)}>0$ and ${\widetilde{A}}^{(2)}>0$. Therefore we need to solve the two branches  of the TBA equations  (\ref{TBA-F}). Without losing generality, in order to capture the main features of the GPs at low temperatures, here we only consider the leading order of the effective pressures.  By iteration the TBA equations, the pressures of excess fermions and bound pairs are respectively given by 
\begin{eqnarray}\label{p}
  {\widetilde{p}}^{(1)} &=& -\frac{1}{2\sqrt{2\pi}}t^{\frac{3}{2}} \text{Li}_{\frac{3}{2}}(-\mathrm{e}^{{\widetilde{A}}^{(1)}/t}),\nonumber\\
  {\widetilde{p}}^{(2)} &=& -\frac{1}{2\sqrt{\pi}}t^{\frac{3}{2}} \text{Li}_{\frac{3}{2}}(-\mathrm{e}^{{\widetilde{A}}^{(2)}/t}),\nonumber\\
  {\widetilde{A}}^{(1)} &=& {\widetilde{\mu}}_1-2{\widetilde{p}}^{(2)},\nonumber\\
  {\widetilde{A}}^{(2)} &=& {\widetilde{\mu}}_2-{\widetilde{p}}^{(2)}-4{\widetilde{p}}^{(1)}.
\end{eqnarray}
Then we  obtain  the leading terms of the  derivatives of the pressure
\begin{eqnarray}
 \frac{\partial \widetilde{p}}{\partial \widetilde \mu}&=&\frac{1}{\sqrt{2}\pi}{{\widetilde{\mu}}_1}^\frac{1}{2}+\frac{2}{\pi}{{\widetilde{\mu}}_2}^\frac{1}{2},\nonumber\\
   \frac{\partial \widetilde{p}}{\partial t}&=&t (\frac{\pi}{6\sqrt{2}}{{\widetilde{\mu}}_1}^{-\frac{1}{2}}
  +\frac{\pi}{6}{{\widetilde{\mu}}_2}^{-\frac{1}{2}}),\nonumber\\
 \frac{\partial^2 \widetilde{p}}{\partial {\widetilde \mu}^2}&=&
  \frac{1}{2\sqrt{2}\pi}{{\widetilde{\mu}}_1}^{-\frac{1}{2}}
  +\frac{2}{\pi}{{\widetilde{\mu}}_2}^{-\frac{1}{2}},\nonumber\\
  \frac{\partial^2 \widetilde{p}}{\partial t^2}&=&\frac{\pi}{6\sqrt{2}}{{\widetilde{\mu}}_1}^{-\frac{1}{2}}
  +\frac{\pi}{6}{{\widetilde{\mu}}_2}^{-\frac{1}{2}}, \nonumber\\
  \frac{\partial^2 \widetilde{p}}{\partial \widetilde \mu \partial t}&=&t(
  -\frac{\pi}{12\sqrt{2}}{{\widetilde{\mu}}_1}^{-\frac{3}{2}}
  -\frac{\pi}{6}{{\widetilde{\mu}}_2}^{-\frac{3}{2}}), \nonumber\\
     \frac{\partial^2 \widetilde{p}}{\partial h\partial  t}&=&-\frac{\pi}{24\sqrt{2}}{{\widetilde{\mu}}_1}^{-\frac{3}{2}}t, \nonumber\\
  \frac{\partial^2 \widetilde{p}}{\partial {\widetilde \mu}\partial  h}&=&\frac{1}{4\sqrt{2}\pi}{{\widetilde{\mu}}_1}^{-\frac{1}{2}}. 
\end{eqnarray}
Thus by definition Eq.(\ref{GCE_GR}),  the Gr\"uneisen parameter for  the FFLO-like phase is  given by 
\begin{equation}\label{gr_FFLO}
	\Gamma=2+\frac{\frac{1}{\sqrt{2} }\left(\lambda -\frac{2}{\lambda}\right) \left(\lambda-\frac{1}{2\lambda}  \right)}{\frac{1}{4\lambda}+2\lambda+\frac{5}{2\sqrt{2}}},
\end{equation}
where the $\lambda$ is defined by $\lambda=\sqrt{{\widetilde{\mu}}_1/{\widetilde{\mu}}_2}$ which is related to the ratio between the densities  of unpaired fermions and bound pairs, i.e. ${\widetilde \mu}_1/{\widetilde \mu}_2 \sim n_1/n_2$. This expression is valid only for strong coupling limit that the rescaled temperature $t $ is much less than unity. 
Here the effective chemical potentials of   unpaired fermions and paired fermions are given by 
$  {\widetilde{\mu}}_1=\widetilde{\mu}+\frac{1}{2}h$ and 
 $ {\widetilde{\mu}}_2=2\widetilde{\mu}+1$, respectively. 
In the Fig.~\ref{Fig3}(a), we observe a good agreement between  the numerical result obtained from the TBA equations (\ref{TBA-F})  and  the  analytical results Eq.(\ref{gr_FFLO}) at  different effective chemical potentials.  For small and large values of the $\lambda$, the GP increases due to approaching to the phase boundaries.

Similarly, at low temperatures, $t\ll 1$, we  obtain the magnetic Gr\"uneisen parameter  from  Eq.(\ref{GCE_GRmag})
\begin{equation}\label{gr_magFFLO}
	\Gamma_{\mathrm{mag}}=-\frac{h}{{\widetilde{\mu}}_2}\frac{1-\frac{2}{\lambda^2}}{8\sqrt{2}\lambda+\frac{\sqrt{2}}{\lambda}+10}. 
\end{equation}
Here  $h=2\tilde{\mu}_2\lambda^2 -\tilde{\mu}_2+1$.
The magnetic GP is also a dimensionless parameter and related to the magnetocaloric effect, see Fig.~\ref{Fig3}.  
In this figure, we also observe a good agreement between numerical and analytical results of the GPS at low temperature  $T=1.0\times 10^{-5}$ in the strong coupling $|c|=10$ and for  different total chemical potentials. 
From the Eq.(\ref{gr_magFFLO}), we  observe that the magnetic GP is equal to zero when the $\lambda^2=2$. 
This point corresponds to the effective chemical potentials of excess fermions and bound pairs are equal, named a critical polarization. 
%

Moreover, with the help of the relation Eq.(\ref{eq:identity}), we can further obtain the interaction GP through  the relation  $\Gamma_{\mathrm{int}}=2-\Gamma-2\Gamma_{\mathrm{mag}}$. Substituting Eq.(\ref{gr_FFLO}) and Eq.(\ref{gr_magFFLO}) into Eq.(\ref{eq:identity}), we have 
\begin{equation}\label{gr_intFFLO}
	\Gamma_{int}=\frac{\left(\frac{1}{{\widetilde{\mu}}_2}-\lambda^2\right)\left(1-\frac{2}{\lambda^2}\right)}{\frac{1}{\sqrt{2}\lambda}+4\sqrt{2}\lambda+5}.
\end{equation} 
In the Fig.~\ref{Fig3}(c), we observe that  the result Eq.(\ref{gr_intFFLO}) is in  a good agreement with the numerical result obtained from the TBA equations  (\ref{TBA-F}) according to the definition Eq.(\ref{GCE_GRint}). 
This agreement  further confirms the identity Eq.(\ref{eq:identity}) among the three types of GPs. 
The interaction GP shows an interacting caloric effect which looks opposite to the magnetocaloric effect. 
We also notice  the  zero point of the interacting GP at $\lambda^2 =2$,  see  Fig.~\ref{Fig3}(c). 
This point is very special because the GP is a constant $\Gamma =2$, i.e. the energy spectrum has a scaling invariant.
%
This opens  a new way to realize adiabatic interacting   quantum cooling in quantum gases of ultracold atoms, also see cooling and thermometry of atomic Fermi gases \cite{Onofrio:2016}

\begin{figure}[]
\begin{center}
\includegraphics[width=0.9\linewidth]{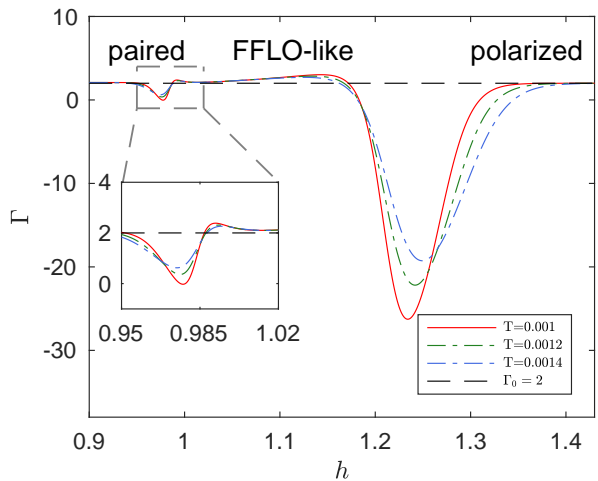}
\end{center}
\caption{The Gr\"uneisen parameter vs. magnetic field for fixed rescaled chemical potential ${\mu}=-0.245$.  By the definition  (\ref{GCE_GR}), the numerical result of the GP is obtained from the TBA equation  (\ref{TBA-F}).  For the strong coupling region, we observe  that the GP is very close to the line $\Gamma=2$ in the fully-polarized phase and in the fully-paired phase due to scaling invariant nature. It  shows the universal singular behavior near the quantum phase transition.}
\label{Fig4}
\end{figure}

\begin{figure}[]
\begin{center}
\includegraphics[width=0.9\linewidth]{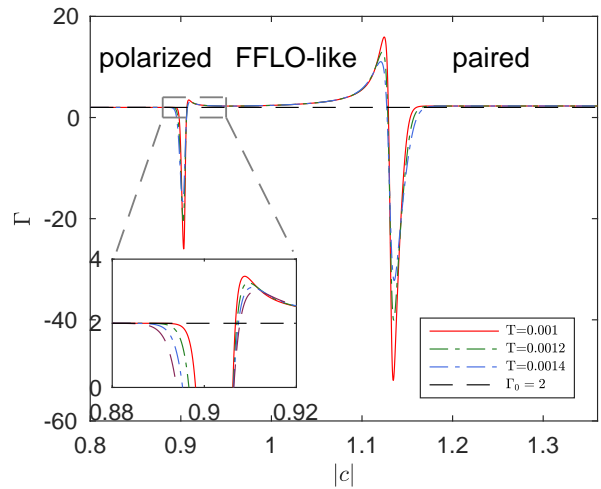}
\end{center}
\caption{The Gr\"uneisen parameter vs. interaction strength $c$  for fixed chemical potential  $\mu=-0.2$ and magnetic field $H=0.5$. By the definition  (\ref{GCE_GR}), the numerical result of the GP is obtained from the TBA equation  (\ref{TBA-F}).  For the strong coupling region, we observe  that like Fig.\ref{Fig4}, the GP is very close to the line $\Gamma=2$ in the fully-polarized phase and in the fully-paired phase due to scaling invariant nature. It  shows the universal singular behavior near the quantum phase transition. }\label{Fig5}
\end{figure}

\subsection{III.4. The Gr\"{u}neisen parameters at  quantum criticality  }

Quantum phase transitions occur  in the 1D attractive Fermi gas  at zero temperature by varying the  external fields like chemical potential and magnetic field. 
In the Fig.\ref{Fig2},  we find  that  the Gr\"uneisen parameters display  singular  behavior near the quantum critical points. 
Such a kind of  singular behaviour has been used to signify quantum phase transition \cite{Gegenwart_RPP_2016,Souza_EJP_2016}.
Hence, the GPs can be regarded as good probe for  quantum criticalities and universal scaling behaviors. 
An important observation of the GPs is that both the magnetic field and interacting GPs characterize the caloric effects for quantum cooling and heat engines. 
Here we further study the GPs for the 1D attractive Yang-Gaudin model, in which the potentials drive the system into four different phases, i.e.  vacuum, fully paired, FFLO-like pairing (partially polarized) and fully polarized phases, see the Fig.~\ref{Fig2} (a). 

Fig.\ref{Fig4} and Fig.\ref{Fig5} show the magnetic field and interacting  GPs at different  temperatures, respectively. 
In Fig.\ref{Fig4}, we fixed the chemical potential ${\mu}=-0.245$.  The long black dashed line stands  for the  constant value $\Gamma =2$ in the two figures. 
We observe that GPs  are  very close to the value of $\Gamma =2$  in the fully polarized phase and in the fully paired phase, where for $t\ll 1$ the system has scaling invariant nature.
However,  near  the phase boundaries,  they are suddenly enhanced and show a universal divergent  nature. 
This elegantly  characterizes  the critical phenomenon of  the model, i.e.  thermodynamic quantities display  universal quantum scalings in the vicinities of critical points as temperature approaches to zero. 
What follows will discuss the scaling functions of the GPs  near  the two non-trivial phase transitions:  from fully-paired phase to the FFLO like phase and from FFLO-like state to the fully-polarized phase. The calculation is presented in the Appendix.

\begin{figure}[]
\begin{center}
\includegraphics[width=0.9\linewidth]{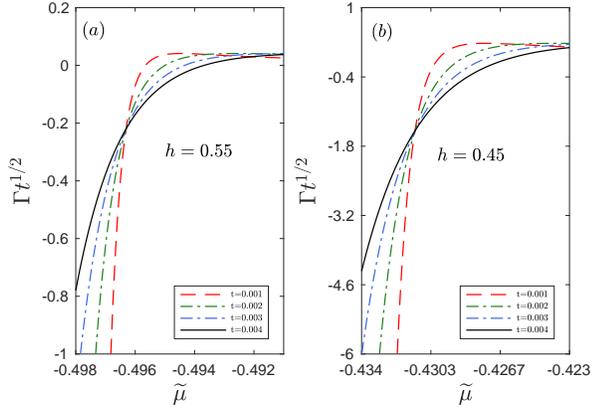}
\end{center}
\caption{Universal scaling behavior of the GP vs the rescaled chemical potential $\tilde{\mu}$ at different temperatures $t=0.001$, $0.002$, $0.003$, $0.004$. The GP is calculated by using the TBA equation  (\ref{TBA-F}). The intersections in the left and right panels give the phase boundaries of the $\mathrm{F-FFLO}$ and $\mathrm{P-FFLO}$ transitions, respectively.    }
\label{Fig6}
\end{figure}

{\bf Phase transition $\mathrm{P-FFLO}$.} Firstly, we discuss the GPs near the phase boundary between the fully paired phase and the FFLO like state. 
We assume that the phase transition from the fully paired phase to  the FFLO like phase is driven by the effective chemical potential ${\widetilde{\mu}}$. The phase boundary  $\mathrm{P-FFLO}$ for strong coupling is determined by  ${\widetilde{\mu}}_{c1}=-\frac{h}{2}+\frac{4}{3\pi}(1-h)^{\frac{3}{2}}$, where $0<1-h \ll 1$.
Up to the second order of  $(1-h)$, we get the scaling functions  of the thermodynamics quantities:
 \begin{eqnarray}\label{P-FFLO}
  \widetilde{n}=\frac{\partial{\widetilde{p}}}{\partial {\widetilde{\mu}}}&=&\frac{2}{\pi}\sqrt{b}-\frac{1}{2\sqrt{2\pi}}t^{\frac{1}{2}}\mathcal{R}_0(\frac{{\widetilde{A}}_1}{t})\nonumber\\
  \widetilde{s}=\frac{\partial{\widetilde{p}}}{\partial {t}}
  &=&\frac{1}{2\sqrt{2\pi}}t^{\frac{1}{2}}\mathcal{R}_1(\frac{{\widetilde{A}}_1}{t})\nonumber\\
  \widetilde{m}=\frac{\partial^2{\widetilde{p}}}{\partial{\widetilde{\mu}}^2}&=&\frac{2}{\pi\sqrt{b}}+\frac{1}{2\sqrt{2\pi}}t^{-\frac{1}{2}}\mathcal{R}_2(\frac{{\widetilde{A}}_1}{t})\nonumber\\
  \widetilde{c_v}/t=\frac{\partial^2{\widetilde{p}}}{\partial{t}^2}&=&-\frac{1}{2\sqrt{2\pi}}t^{-\frac{1}{2}}\mathcal{R}_3(\frac{{\widetilde{A}}_1}{t})\nonumber\\
  \frac{\partial^2{\widetilde{p}}}{\partial{\widetilde{\mu}}\partial t}&=&
  \frac{1}{2\sqrt{2\pi}}t^{-\frac{1}{2}}\mathcal{R}_4(\frac{{\widetilde{A}}_1}{t})\nonumber\\
    \frac{\partial^2{\widetilde{p}}}{\partial{\widetilde{h}}\partial t}&=&
  \frac{1}{4\sqrt{2\pi}}t^{-\frac{1}{2}}\mathcal{R}_4(\frac{{\widetilde{A}}_1}{t})\nonumber\\
  \frac{\partial^2{\widetilde{p}}}{\partial{\widetilde{\mu}}\partial h}&=&
  \frac{1}{4\sqrt{2\pi}}t^{-\frac{1}{2}}\mathcal{R}_2(\frac{{\widetilde{A}}_1}{t}), 
  \end{eqnarray}
  here  $\widetilde{A}_1=(\widetilde{\mu}-\widetilde{\mu}_{c1})$ and $b=(1-h)(1+\frac{2}{\pi}\sqrt{1-h})$.
 While the analytic function $\mathcal{R}_i(i=0,1,2,3,4)$ are given by 
\begin{eqnarray}
  \mathcal{R}_0(x)&=&\text{Li}_{\frac{1}{2}}(-\mathrm{e}^x)\nonumber\\
  \mathcal{R}_1(x)&=&-\frac{3}{2}\text{Li}_{\frac{3}{2}}(-\mathrm{e}^x)+x\text{Li}_{\frac{1}{2}}(-\mathrm{e}^x)\nonumber\\
  \mathcal{R}_2(x)&=&-\text{Li}_{-\frac{1}{2}}(-\mathrm{e}^x)\nonumber\\
  \mathcal{R}_3(x)&=&\frac{3}{4}\text{Li}_{\frac{3}{2}}(-\mathrm{e}^x)-x\text{Li}_{\frac{1}{2}}(-\mathrm{e}^x)+x^2\text{Li}_{-\frac{1}{2}}(-\mathrm{e}^x)\nonumber\\
  \mathcal{R}_4(x)&=&-\frac{1}{2}\text{Li}_{\frac{1}{2}}(-\mathrm{e}^x)+x\text{Li}_{-\frac{1}{2}}(-\mathrm{e}^x).
\end{eqnarray}
By  the definition of the GP  (\ref{GCE_GR}), we obtain the an explicit form 
 \begin{eqnarray}
 \Gamma &=&\frac{4\sqrt{2b}}{\pi}t^{-\frac{1}{2}}\mathcal{G}(\frac{\widetilde{\mu}-\widetilde{\mu}_{c1}}{t}),
 \end{eqnarray}
  where $b=(1-h)(1+\frac{2}{\pi}\sqrt{1-h})$ and
    \begin{equation}\label{G-function}
   \mathcal{G}(x)=\frac{\mathcal{R}_4}{\mathcal{R}_2\mathcal{R}_3+\mathcal{R}_4\mathcal{R}_4}.
 \end{equation}
 As the rescaled temperature $t\ll 1$, the GP  (\ref{GCE_GR}) has an universal scaling $t^{-\frac{1}{2}}$  for the phase transition $\mathrm{P-FFLO}$. 
 In  Fig.\ref{Fig6}(b),  we demonstrate this universal scaling behavior for the phase transition $\mathrm{P-FFLO}$.  

Similarly, by the definition of the magnetic GP  Eq. (\ref{GCE_GRmag}),
we  obtain the scaling function of the magnetic GP 
 \begin{eqnarray}
 \Gamma_{\mathrm{mag}} &=&h \frac{2\sqrt{2}}{\sqrt{\pi}\sqrt{b}}t^{-\frac{1}{2}}\mathcal{G}(\frac{\widetilde{\mu}-\widetilde{\mu}_{c1}}{t}).
 \end{eqnarray}
 This shows a similar universal scaling behaviour of the magnetic GP, $\Gamma_{\rm mag} \propto t^{-\frac{1}{2}}$.

Regarding the scaling property of the  pressure $p=\frac{c^3}{2}\widetilde{p}(\frac{\mu}{c^2/2},\frac{T}{c^2/2},\frac{H}{c^2/2})$, we can obtain
\begin{eqnarray}
\frac{\partial^2 {p}}{\partial c \partial T}&=&\frac{1}{c}[\frac{\partial {p}}{\partial T}-2\mu\frac{\partial^2{p}}{\partial{\mu}\partial T}-2T\frac{\partial^2{p}}{\partial T^2}-2H\frac{\partial^2{p}}{\partial H \partial T}]\nonumber\\
\frac{\partial^2 {p}}{\partial c \partial \mu}&=&\frac{1}{c}[\frac{\partial {p}}{\partial \mu}-2\mu\frac{\partial^2{p}}{\partial{\mu}^2}-2T\frac{\partial^2{\widetilde{p}}}{\partial T \partial \mu}-2H\frac{\partial^2{p}}{\partial H \partial \mu}]. 
\end{eqnarray}
we have the following results 
\begin{eqnarray}
\frac{\partial^2 {p}}{\partial c \partial T}&=&\frac{\sqrt{2}}{c}(-\mu\frac{1}{\sqrt{2\pi}}-H\frac{1}{2\sqrt{2\pi}})T^{-\frac{1}{2}}\mathcal{R}_4(\frac{{{A}}_1}{T})\nonumber\\
\frac{\partial^2 {p}}{\partial c \partial \mu}&=&\frac{2}{\pi}\sqrt{b}-\frac{c^2}{2}\mu\frac{4}{\pi \sqrt{b}}-\frac{\sqrt{2}}{c}\mu\frac{1}{\sqrt{2\pi}}T^{-\frac{1}{2}}\mathcal{R}_2(\frac{{{A}}_1}{T})\nonumber\\
&& -\frac{\sqrt{2}}{c}H\frac{1}{2\sqrt{2\pi}}T^{-\frac{1}{2}}\mathcal{R}_2(\frac{{{A}}_1}{T})]. 
\end{eqnarray}
From the  definition of the interaction GP (\ref{GCE_GRint}),
we have an explicit form of scaling function
 \begin{eqnarray}
 \Gamma_{\mathrm{int}} &=&-\frac{4\sqrt{2}\sqrt{b}}{\sqrt{\pi}}t^{-\frac{1}{2}}\mathcal{G}(\frac{\widetilde{\mu} -\widetilde{\mu}_{c1}}{t})\nonumber\\
 &&-h\frac{4\sqrt{2}}{\sqrt{\pi}\sqrt{b}}t^{-\frac{1}{2}}\mathcal{G}(\frac{\widetilde{\mu}-\widetilde{\mu}_{c1}}{t})
 \end{eqnarray}
It can be seen from the scaling functions that  three types of GPs satisfy the identity
\begin{eqnarray}\label{gr_relation}
\Gamma=2-2\Gamma_{\mathrm{mag}}-\Gamma_{\mathrm{int}}. 
\end{eqnarray}

{\bf Phase transition $\mathrm{FFLO-F}$.}
We now consider the GPs near the phase transition from  the  FFLO-like pairing phase to the fully polarized phase. 
This  phase transition occurs by the variation of the  chemical potential ${\widetilde{\mu}}$ too. 
The critical phase boundary is determined by ${\widetilde{\mu}}_{c2}=-\frac{1}{2}+\frac{1}{3\pi}(h-1)^{\frac{3}{2}}$, here $h>1$. We expand it to the second order of  $(h-1)$,  we obtained  the scaling functions of  the thermodynamics quantities:
\begin{eqnarray}
  \frac{\partial{\widetilde{p}}}{\partial {\widetilde{\mu}}}&=&\frac{\sqrt{a}}{2\pi}-\frac{1}{\sqrt{\pi}}t^{\frac{1}{2}}\mathcal{R}_0(\frac{{\widetilde{A}}_2}{t}),\nonumber\\
  \frac{\partial{\widetilde{p}}}{\partial {t}}
  &=&\frac{1}{2\sqrt{\pi}}t^{\frac{1}{2}}\mathcal{R}_1(\frac{{\widetilde{A}}_2}{t}), \nonumber\\
  \frac{\partial^2{\widetilde{p}}}{\partial{\widetilde{\mu}}^2}&=&\frac{1}{2\pi\sqrt{a}}+\frac{2}{\sqrt{\pi}}t^{-\frac{1}{2}}\mathcal{R}_2(\frac{{\widetilde{A}}_2}{t}), \nonumber\\
  \frac{\partial^2{\widetilde{p}}}{\partial{t}^2}&=&-\frac{1}{2\sqrt{\pi}}t^{-\frac{1}{2}}\mathcal{R}_3(\frac{{\widetilde{A}}_2}{t}), \nonumber\\
  \frac{\partial^2{\widetilde{p}}}{\partial{\widetilde{\mu}}\partial t}&=&
  \frac{1}{\sqrt{\pi}}t^{-\frac{1}{2}}\mathcal{R}_4(\frac{{\widetilde{A}}_2}{t}), \nonumber\\
    \frac{\partial^2{\widetilde{p}}}{\partial{\widetilde{h}}\partial t}&=&
  -\frac{\sqrt{a}}{2{\pi}^{\frac{3}{2}}}t^{-\frac{1}{2}}\mathcal{R}_4(\frac{{\widetilde{A}}_2}{t}), \nonumber\\
  \frac{\partial^2{\widetilde{p}}}{\partial{\widetilde{\mu}}\partial h}&=&
  \frac{1}{4\pi}\frac{1}{\sqrt{a}}-\frac{\sqrt{a}}{{\pi}^{\frac{3}{2}}}t^{-\frac{1}{2}}\mathcal{R}_2(\frac{{\widetilde{A}}_1}{t}),
\end{eqnarray}
 where by definition $\widetilde{A}_2=2(\widetilde{\mu}-\widetilde{\mu}_{c2})$ and $\widetilde{\mu}_{c2}$ denotes the critical chemical potential.
In the above equation  $a=(h-1)(1+\frac{2}{3\pi}\sqrt{h-1})$, which  is almost a constant.  The scaling function of the GP at this phase boundary is given by 
 \begin{eqnarray}
   \Gamma &=&\frac{\sqrt{a}}{2\sqrt{\pi}}t^{-\frac{1}{2}}\mathcal{G}(\frac{2({\widetilde{\mu}}-{\widetilde{\mu}}_{c2})}{t}). 
 \end{eqnarray}
 The GP near the phase transition shows the same scaling behaviour $ \Gamma \propto t^{-\frac{1}{2}}$ as that for the phase transition $\mathrm{P-FFLO}$.
Moreover, the scaling function of the magnetic GP near the phase transition $\mathrm{FFLO-F}$ is given by 
 \begin{eqnarray}
   \Gamma_{\mathrm{mag}} =-h\frac{\frac{1}{\pi}+\frac{1}{\sqrt{a}}}{ { 4} \sqrt{\pi}}t^{-\frac{1}{2}}\mathcal{G}(\frac{2({\widetilde{\mu}}-{\widetilde{\mu}}_{c2})}{t}).
 \end{eqnarray}
 In order to  calculate  the interacting GP (\ref{GCE_GRint}),  we obtain the quantities 
\begin{eqnarray}
\frac{\partial^2 {p}}{\partial c \partial T}&=&\frac{\sqrt{2}}{c}(-\mu\frac{2}{\sqrt{\pi}}+H\frac{\sqrt{a}}{{\pi}^{\frac{3}{2}}})T^{-\frac{1}{2}}\mathcal{R}_4(\frac{{{A}}_1}{T}),\nonumber\\
\frac{\partial^2 {{p}}}{\partial c \partial {\mu}}&=&-\frac{2}{c^2}{\mu}\frac{1}{\pi \sqrt{a}}-\frac{2}{c^2}H\frac{1}{2\pi \sqrt{a}}-\frac{\sqrt{2}}{c}\mu\frac{4}{\sqrt{\pi}}T^{-\frac{1}{2}}\mathcal{R}_2(\frac{{{A}}_1}{T})\nonumber\\
&&+\frac{\sqrt{2}}{c}H\frac{2\sqrt{a}}{{\pi}^{\frac{3}{2}}}T^{-\frac{1}{2}}\mathcal{R}_2(\frac{{{A}}_1}{T})].
\end{eqnarray}
 Then  we obtain  the scaling function of the interacting GP
  \begin{eqnarray}
   \Gamma_{\mathrm{int}} &=& h\frac{\frac{1}{\pi}+\frac{1}{\sqrt{a}}}{2\sqrt{\pi}}t^{-\frac{1}{2}}\mathcal{G}(\frac{2({\widetilde{\mu}}-{\widetilde{\mu}}_{c2})}{t})\nonumber\\
   &&-\frac{\sqrt{a}}{2\sqrt{\pi}}t^{-\frac{1}{2}}\mathcal{G}(\frac{2({\widetilde{\mu}}-{\widetilde{\mu}}_{c2})}{t}).
 \end{eqnarray}
 Fig.~\ref{Fig6}(a) shows  the scaling behavior  near the phase transition $ \mathrm{FFLO-F}$. 
 We further prove that  these scaling functions of the GP at quantum criticality also satisfy  the identity
\begin{eqnarray}
\Gamma=2-2\Gamma_{\mathrm{mag}}-\Gamma_{\mathrm{int}}. 
\end{eqnarray}

\section{IV. Conclusion}
\label{sectionV}
We  have derived    various expressions of the Gr\"{u}neisen parameters  for  studying magneto- and interaction-driven-caloric  effects of quantum gases.
Using the Bethe ansatz solution, we  have obtained the  expansionary, magnetic and interacting Gr\"uneisen parameters   for  various quantum phases in the 1D integrable Bose  and Fermi gases.
The obtained Gr\"uneisen parameters  confirm the  identity  found in  \cite{Yu:2019}, revealing an important symmetry of the thermal potentials. 
These GPs elegantly  quantify   the dependences of  characteristic energy scales   of   quantum gases on the  volume, the magnetic field and the interaction strength, and present the caloric effects induced by the variations  of these potentials. 
In particular, such different  GPs for  the 1D attractive Fermi gas significantly quantify the magneto- and interaction-driven-caloric  effects of quantum gases in the fully-paired, FFLO like and fully-polarized phases. 
We have also obtained  universal scaling behaviors of the GPs in the vicinities of  the quantum critical points in the interacting Fermi gas. 
It turns out that the divergence of the GPs  not only  provides an experimental identification of non-Fermi liquid nature at quantum criticality but also remarkably  determine full phase diagram of the gases at low temperature regimes. 
In addition, our results provide   an alternative way of quantum cooling through the  quantum adiabatic interaction ramp-up and -down in quantum gases of ultracold atoms. 
Our methods opens to further study the  interaction- and magnetic-field-driven quantum refrigeration  and  quantum heat engine in quantum gases of ultracold atoms.
%



\clearpage

\begin{widetext}

\appendix
\section{Appendix }
1.density.
\begin{eqnarray*}
\widetilde{n}&=&\frac{\partial{\widetilde{p}}}{\partial{\widetilde{\mu}}} \\
&=&-\frac{1}{2\sqrt{2\pi}}F^{(1)}_{\frac{1}{2}}-\frac{1}{\sqrt{\pi}}F^{(2)}_{%
\frac{1}{2}}-\frac{\sqrt{2}}{\pi}F^{(1)}_{\frac{1}{2}}F^{(2)}_{\frac{1}{2}} -%
\frac{1}{2\pi}(F^{(2)}_{\frac{1}{2}})^2-\frac{1}{2\pi^{\frac{3}{2}}}%
(F^{(1)}_{\frac{1}{2}})^2F^{(2)}_{\frac{1}{2}} \\
&-&\frac{3}{\sqrt{2}\pi^{\frac{3}{2}}}F^{(1)}_{\frac{1}{2}}(F^{(2)}_{\frac{1%
}{2}})^2 -\frac{1}{4\pi^{\frac{3}{2}}}(F^{(2)}_{\frac{1}{2}})^3
\end{eqnarray*}

2.magnetization.
\begin{eqnarray*}
\widetilde{m}&=&\frac{\partial{\widetilde{p}}}{\partial{h}} \\
&=&-\frac{1}{4\sqrt{2\pi}}F^{(1)}_{\frac{1}{2}}-\frac{1}{2\sqrt{2}\pi}%
F^{(1)}_{\frac{1}{2}}F^{(2)}_{\frac{1}{2}} -\frac{1}{4\pi^{\frac{3}{2}}}%
F^{(1)}_{\frac{1}{2}}F^{(1)}_{\frac{1}{2}}F^{(2)}_{\frac{1}{2}} -\frac{1}{4%
\sqrt{2}{\pi}^{\frac{3}{2}}}F^{(1)}_{\frac{1}{2}}F^{(2)}_{\frac{1}{2}%
}F^{(2)}_{\frac{1}{2}}
\end{eqnarray*}

3.susceptibility.
\begin{eqnarray*}
\widetilde{\chi}&=&\frac{\partial^2\widetilde{p}}{\partial h^2} \\
&=&-\frac{1}{8\sqrt{2\pi}}F^{(1)}_{-\frac{1}{2}}-\frac{1}{4\sqrt{2}\pi}%
F^{(1)}_{-\frac{1}{2}}F^{(2)}_{\frac{1}{2}} -\frac{1}{4\pi^{\frac{3}{2}}}%
F^{(1)}_{\frac{1}{2}}F^{(1)}_{\frac{1}{2}}F^{(2)}_{-\frac{1}{2}} -\frac{3}{%
8\pi^{\frac{3}{2}}}F^{(1)}_{-\frac{1}{2}}F^{(1)}_{\frac{1}{2}}F^{(2)}_{\frac{%
1}{2}}-\frac{1}{8\sqrt{2}\pi^{\frac{3}{2}}}F^{(1)}_{-\frac{1}{2}}F^{(2)}_{\frac{1%
}{2}}F^{(2)}_{\frac{1}{2}}
\end{eqnarray*}

4.compressibility.
\begin{eqnarray*}
\widetilde{\kappa}&=&\frac{\partial^2\widetilde{p}}{\partial \widetilde{\mu}^2} \\
&=&-\frac{1}{2\sqrt{2\pi}}F^{(1)}_{-\frac{1}{2}}-\frac{2}{\sqrt{\pi}}%
F^{(2)}_{-\frac{1}{2}}-\frac{3\sqrt{2}}{\pi}F^{(1)}_{\frac{1}{2}}F^{(2)}_{-\frac{1}{2}} -\frac{3%
}{\sqrt{2}\pi}F^{(1)}_{-\frac{1}{2}}F^{(2)}_{\frac{1}{2}}-\frac{3}{\pi}%
F^{(2)}_{-\frac{1}{2}}F^{(2)}_{\frac{1}{2}} \\
&-&\frac{3}{\pi^{\frac{3}{2}}}F^{(1)}_{\frac{1}{2}}F^{(1)}_{\frac{1}{2}%
}F^{(2)}_{-\frac{1}{2}} -\frac{3}{2\pi^{\frac{3}{2}}}F^{(1)}_{-\frac{1}{2}%
}F^{(1)}_{\frac{1}{2}}F^{(2)}_{\frac{1}{2}} -\frac{21}{\sqrt{2}\pi^{\frac{3}{%
2}}}F^{(1)}_{\frac{1}{2}}F^{(2)}_{-\frac{1}{2}}F^{(2)}_{\frac{1}{2}}-\frac{15}{2\sqrt{2}\pi^{\frac{3}{2}}}F^{(1)}_{-\frac{1}{2}}F^{(2)}_{\frac{%
1}{2}}F^{(2)}_{\frac{1}{2}} -\frac{3}{\pi^{\frac{3}{2}}}F^{(2)}_{-\frac{1}{2}%
}F^{(2)}_{\frac{1}{2}}F^{(2)}_{\frac{1}{2}}
\end{eqnarray*}

5.entropy
\begin{eqnarray*}
\widetilde{s}&=&\frac{\partial{\widetilde{p}}}{\partial{t}} \\
&=&-\frac{3}{4\sqrt{2\pi}t}F^{(1)}_{\frac{3}{2}}-\frac{3}{4\sqrt{\pi}t}%
F^{(2)}_{\frac{3}{2}} +\frac{\widetilde{A}^{(1)}}{2\sqrt{2\pi}t}F^{(1)}_{%
\frac{1}{2}} +\frac{\widetilde{A}^{(2)}}{2\sqrt{\pi}t}F^{(2)}_{\frac{1}{2}}
-\frac{3}{2\sqrt{2}\pi t}F^{(1)}_{\frac{3}{2}}F^{(2)}_{\frac{1}{2}} -\frac{%
3}{4\sqrt{2}\pi t}F^{(1)}_{\frac{1}{2}}F^{(2)}_{\frac{3}{2}}\\
 &-&\frac{3}{8\pi t}F^{(2)}_{\frac{1}{2}}F^{(2)}_{\frac{3}{2}}
+\frac{\widetilde{A}^{(1)}}{\sqrt{2}\pi t}F^{(1)}_{\frac{1}{2}}F^{(2)}_{%
\frac{1}{2}} +\frac{\widetilde{A}^{(2)}}{2\sqrt{2}\pi t}F^{(1)}_{\frac{1}{2}%
}F^{(2)}_{\frac{1}{2}} +\frac{\widetilde{A}^{(2)}}{4\pi t}F^{(2)}_{\frac{1}{2%
}}F^{(2)}_{\frac{1}{2}} \\
\end{eqnarray*}

6.specific heat
\begin{eqnarray*}
\widetilde{c}_V/t &=& \frac{\partial^2\widetilde{p}}{\partial t^2} \\
&=& -\frac{(\widetilde{A}^{(1)})^2}{2\sqrt{2\pi}t^2}F^{(1)}_{-\frac{1}{2}} +%
\frac{\widetilde{A}^{(1)}}{2\sqrt{2\pi}t^2}F^{(1)}_{\frac{1}{2}} -\frac{3}{8%
\sqrt{2\pi}t^2}F^{(1)}_{\frac{3}{2}} -\frac{(\widetilde{A}^{(2)})^2}{2\sqrt{%
\pi}t^2}F^{(2)}_{-\frac{1}{2}} +\frac{\widetilde{A}^{(2)}}{2\sqrt{\pi}t^2}%
F^{(2)}_{\frac{1}{2}} -\frac{3}{8\sqrt{\pi}t^2}F^{(2)}_{\frac{3}{2}} \\
&-&\frac{\sqrt{2}\widetilde{A}^{(1)}\widetilde{A}^{(2)}}{\pi t^2}F^{(1)}_{%
\frac{1}{2}}F^{(2)}_{-\frac{1}{2}} -\frac{(\widetilde{A}^{(2)})^2}{2\sqrt{2}%
\pi t^2}F^{(1)}_{\frac{1}{2}}F^{(2)}_{-\frac{1}{2}} +\frac{3\widetilde{A}%
^{(2)}}{\sqrt{2}\pi t^2}F^{(1)}_{\frac{3}{2}}F^{(2)}_{-\frac{1}{2}} -\frac{({%
\widetilde{A}}^{(1)})^2}{\sqrt{2}\pi t^2}F^{(1)}_{-\frac{1}{2}}F^{(2)}_{%
\frac{1}{2}} \\
&-&\frac{\widetilde{A}^{(1)}\widetilde{A}^{(2)}}{\sqrt{2}\pi t^2}F^{(1)}_{-%
\frac{1}{2}}F^{(2)}_{\frac{1}{2}} +\frac{\sqrt{2}\widetilde{A}^{(1)}}{\pi t^2%
}F^{(1)}_{\frac{1}{2}}F^{(2)}_{\frac{1}{2}} +\frac{\widetilde{A}^{(2)}}{%
\sqrt{2}\pi t^2}F^{(1)}_{\frac{1}{2}}F^{(2)}_{\frac{1}{2}} -\frac{9}{4\sqrt{2%
}\pi t^2}F^{(1)}_{\frac{3}{2}}F^{(2)}_{\frac{1}{2}} \\
&-&\frac{3(\widetilde{A}^{(2)})^2}{4\pi t^2}F^{(2)}_{-\frac{1}{2}}F^{(2)}_{%
\frac{1}{2}} +\frac{\widetilde{A}^{(2)}}{2\pi t^2}F^{(2)}_{\frac{1}{2}%
}F^{(2)}_{\frac{3}{2}} +\frac{3\widetilde{A}^{(1)}}{2\sqrt{2}\pi t^2}%
F^{(1)}_{-\frac{1}{2}}F^{(2)}_{\frac{3}{2}} -\frac{9}{8\sqrt{2}\pi t^2}%
F^{(1)}_{\frac{1}{2}}F^{(2)}_{\frac{3}{2}} \\
&+&\frac{3\widetilde{A}^{(2)}}{4\pi t^2}F^{(2)}_{-\frac{1}{2}}F^{(2)}_{\frac{%
3}{2}} -\frac{9}{16\pi t^2}F^{(2)}_{\frac{1}{2}}F^{(2)}_{\frac{3}{2}}
\end{eqnarray*}

7.the magnetization effect
\begin{eqnarray*}
\frac{\partial {\widetilde{M}}}{\partial t} &=&\frac{\partial ^{2}{%
\widetilde{p}}}{\partial h\partial t} \\
&=&-\frac{1}{8\sqrt{2\pi }t}F_{\frac{1}{2}}^{(1)}+\frac{1}{4\sqrt{2\pi }}F_{-%
\frac{1}{2}}^{(1)}\frac{\widetilde{A}^{(1)}}{t} \\
&+&\frac{1}{4\sqrt{2}\pi t}F_{-\frac{1}{2}}^{(1)}F_{\frac{1}{2}}^{(2)}%
\widetilde{A}^{(2)}-\frac{1}{8\pi t}F_{\frac{1}{2}}^{(1)}F_{\frac{1%
}{2}}^{(1)}+\frac{1}{4\sqrt{2}\pi t}F_{\frac{1}{2}}^{(1)}F_{-\frac{1}{2}}^{(2)}\widetilde{A}^{(2)}+\frac{1}{2\sqrt{2}\pi t}F_{-\frac{1}{2}}^{(1)}F_{\frac{1}{2}}^{(2)}%
\widetilde{A}^{(1)} \\
\end{eqnarray*}%

8.
\begin{eqnarray*}
\frac{\partial {\widetilde{n}}}{\partial t} &=&\frac{\partial ^{2}{\widetilde{p}}}{\partial \widetilde{\mu}\partial t}\\
&=&\frac{1}{2\sqrt{2\pi}t}\widetilde{A}^{(1)}F_{-\frac{1}{2}}^{(1)}-\frac{1}{4\sqrt{2\pi}t}F_{\frac{1}{2}}^{(1)}+\frac{1}{\sqrt{\pi}t}\widetilde{A}^{(2)}F_{-\frac{1}{2}}^{(2)}-\frac{1}{2\sqrt{\pi}t}F_{\frac{1}{2}}^{(2)}\\
&+&\frac{1}{2\pi t}\widetilde{A}^{(1)}F_{-\frac{1}{2}}^{(1)}F_{\frac{1}{2}}^{(1)}-\frac{1}{4\pi t}F_{\frac{1}{2}}^{(1)}F_{\frac{1}{2}}^{(1)}+\frac{\sqrt{2}}{\pi t}\widetilde{A}^{(1)}F_{\frac{1}{2}}^{(1)}F_{-\frac{1}{2}}^{(2)}+\frac{1}{\sqrt{2}\pi t}\widetilde{A}^{(2)}F_{\frac{1}{2}}^{(1)}F_{-\frac{1}{2}}^{(2)}\\
&-&\frac{3}{\sqrt{2}\pi t}F_{\frac{3}{2}}^{(1)}F_{-\frac{1}{2}}^{(2)}+\frac{1}{\sqrt{2}\pi t}\widetilde{A}^{(1)}F_{-\frac{1}{2}}^{(1)}F_{\frac{1}{2}}^{(2)}+\frac{1}{2\sqrt{2}\pi t}\widetilde{A}^{(2)}F_{-\frac{1}{2}}^{(1)}F_{\frac{1}{2}}^{(2)}-\frac{1}{\sqrt{2}\pi t}F_{\frac{1}{2}}^{(1)}F_{\frac{1}{2}}^{(2)}\\
&+&\frac{5}{2\pi t}\widetilde{A}^{(2)}F_{-\frac{1}{2}}^{(2)}F_{\frac{1}{2}}^{(2)}-\frac{1}{\pi t}F_{\frac{1}{2}}^{(2)}F_{\frac{1}{2}}^{(2)}-\frac{3}{4\sqrt{2}\pi t}F_{-\frac{1}{2}}^{(1)}F_{\frac{3}{2}}^{(2)}-\frac{3}{4\pi t}F_{-\frac{1}{2}}^{(2)}F_{\frac{3}{2}}^{(2)}
\end{eqnarray*}%

9.
\begin{eqnarray*}
\frac{\partial ^{2}{\widetilde{p}}}{\partial \widetilde{\mu}\partial h}
&=&-\frac{1}{4\sqrt{2\pi}}F_{-\frac{1}{2}}^{(1)}-\frac{1}{\sqrt{2}\pi}F_{\frac{1}{2}}^{(1)}F_{-\frac{1}{2}}^{(2)}-\frac{1}{\sqrt{2}\pi}F_{-\frac{1}{2}}^{(1)}F_{\frac{1}{2}}^{(2)}\\
	&=&-\frac{1}{\pi^{\frac{3}{2}}}F_{\frac{1}{2}}^{(1)}F_{\frac{1}{2}}^{(1)}F_{-\frac{1}{2}}^{(2)}-\frac{3}{4\pi^{\frac{3}{2}}}F_{-\frac{1}{2}}^{(1)}F_{\frac{1}{2}}^{(1)}F_{\frac{1}{2}}^{(2)}-\frac{3}{2\sqrt{2}\pi^{\frac{3}{2}}}F_{\frac{1}{2}}^{(1)}F_{-\frac{1}{2}}^{(2)}F_{\frac{1}{2}}^{(2)}-\frac{3}{2\sqrt{2}\pi^{\frac{3}{2}}}F_{-\frac{1}{2}}^{(1)}F_{\frac{1}{2}}^{(2)}F_{\frac{1}{2}}^{(2)}
\end{eqnarray*}%

where $F_{n}^{(r)}=t^{n}Li_{n}(-e^{\widetilde{A}^{(r)}/t})$, $r=1,2$.
\end{widetext}

\end{document}